\documentstyle[12pt]{article}

\catcode`\@=11
\long\def\@makefntext#1{
\protect\noindent \hbox to 3.2pt {\hskip-.9pt
$^{{\ninerm\@thefnmark}}$\hfil}#1\hfill}                

\def\@makefnmark{\hbox to 0pt{$^{\@thefnmark}$\hss}}  

\def\ps@myheadings{\let\@mkboth\@gobbletwo
\def\@oddhead{\hbox{}
\rightmark\hfil\ninerm\thepage}
\def\@oddfoot{}\def\@evenhead{\ninerm\thepage\hfil
\leftmark\hbox{}}\def\@evenfoot{}
\def\sectionmark##1{}\def\subsectionmark##1{}}

\renewcommand{\thefootnote}{\fnsymbol{footnote}}

\newcounter{sectionc}\newcounter{subsectionc}\newcounter{subsubsectionc}
\renewcommand{\section}[1] {\vspace*{0.6cm}\addtocounter{sectionc}{1}
\setcounter{subsectionc}{0}\setcounter{subsubsectionc}{0}\noindent
        {\normalsize\bf\thesectionc. #1}\par\vspace*{0.4cm}}
\renewcommand{\subsection}[1] {\vspace*{0.6cm}\addtocounter{subsectionc}{1}
        \setcounter{subsubsectionc}{0}\noindent
        {\normalsize\it\thesectionc.\thesubsectionc. #1}\par\vspace*{0.4cm}}
\renewcommand{\subsubsection}[1]
{\vspace*{0.6cm}\addtocounter{subsubsectionc}{1}
        \noindent {\normalsize\rm\thesectionc.\thesubsectionc.
        \thesubsubsectionc. #1}\par\vspace*{0.4cm}}

\newcounter{appendixc}
\newcounter{subappendixc}[appendixc]
\newcounter{subsubappendixc}[subappendixc]

\renewcommand{\appendix}[1] {\vspace*{0.6cm}
        \refstepcounter{appendixc}
        \setcounter{figure}{0}
        \setcounter{table}{0}
        \setcounter{equation}{0}
        \renewcommand{\thefigure}{\Alph{appendixc}.\arabic{figure}}
        \renewcommand{\thetable}{\Alph{appendixc}.\arabic{table}}
        \renewcommand{\theappendixc}{\Alph{appendixc}}
        \renewcommand{\theequation}{\Alph{appendixc}.\arabic{equation}}
        \noindent{\bf Appendix \theappendixc #1}\par\vspace*{0.4cm}}

\def\abstracts#1{{
        \centering{\begin{minipage}{12.2truecm}\footnotesize\baselineskip=12pt
\noindent
        \centerline{\footnotesize ABSTRACT}\vspace*{0.3cm}
        \parindent=0pt #1
        \end{minipage}}\par}}

\renewenvironment{thebibliography}[1]
        {\begin{list}{\arabic{enumi}.}
        {\usecounter{enumi}\setlength{\parsep}{0pt}
\setlength{\leftmargin 1.25cm}{\rightmargin 0pt}
         \setlength{\itemsep}{0pt} \settowidth
        {\labelwidth}{#1.}\sloppy}}{\end{list}}

\newcounter{itemlistc}
\newcounter{romanlistc}
\newcounter{alphlistc}
\newcounter{arabiclistc}

\newcommand{\fcaption}[1]{
        \refstepcounter{figure}
        \setbox\@tempboxa = \hbox{\footnotesize Fig.~\thefigure. #1}
        \ifdim \wd\@tempboxa > 6in
           {\begin{center}
        \parbox{6in}{\footnotesize\baselineskip=12pt Fig.~\thefigure. #1}
            \end{center}}
        \else
             {\begin{center}
             {\footnotesize Fig.~\thefigure. #1}
              \end{center}}
        \fi}

\newcommand{\tcaption}[1]{
        \refstepcounter{table}
        \setbox\@tempboxa = \hbox{\footnotesize Table~\thetable. #1}
        \ifdim \wd\@tempboxa > 6in
           {\begin{center}
        \parbox{6in}{\footnotesize\baselineskip=12pt Table~\thetable. #1}
            \end{center}}
        \else
             {\begin{center}
             {\footnotesize Table~\thetable. #1}
              \end{center}}
        \fi}

\def\@citex[#1]#2{\if@filesw\immediate\write\@auxout
        {\string\citation{#2}}\fi
\def\@citea{}\@cite{\@for\@citeb:=#2\do
        {\@citea\def\@citea{,}\@ifundefined
        {b@\@citeb}{{\bf ?}\@warning
        {Citation `\@citeb' on page \thepage \space undefined}}
        {\csname b@\@citeb\endcsname}}}{#1}}

\newif\if@cghi
\def\cite{\@cghitrue\@ifnextchar [{\@tempswatrue
        \@citex}{\@tempswafalse\@citex[]}}
\def\citelow{\@cghifalse\@ifnextchar [{\@tempswatrue
        \@citex}{\@tempswafalse\@citex[]}}
\def\@cite#1#2{{$\null^{#1}$\if@tempswa\typeout
        {IJCGA warning: optional citation argument
        ignored: `#2'} \fi}}

 1
 1
 1

\font\ninerm=cmr9

\input{psfig}

\begin{document}

\centerline{\normalsize\bf QUANTUM SPIN GLASSES IN FINITE DIMENSIONS}

\vspace*{0.6cm}
\centerline{\footnotesize Heiko Rieger}
\baselineskip=13pt
\centerline{\footnotesize\it Institut f\"ur Theoretische Physik,
Universit\"at zu K\"oln, 50937 K\"oln, Germany}
\centerline{\footnotesize\it and HLRZ c/o Forschungszentrum J\"ulich, 52425
J\"ulich, Germany}
\centerline{\footnotesize E-mail: rieger@thp.uni-koeln.de}
\vspace*{0.3cm}
\centerline{\footnotesize and}
\vspace*{0.3cm}
\centerline{\footnotesize A.\ Peter Young}
\baselineskip=13pt
\centerline{\footnotesize\it Physics Department,
University of California at Santa Cruz}
\centerline{\footnotesize\it Santa Cruz, CA 95064, USA}

\vspace*{0.9cm} \abstracts{ The Ising spin glass model in a transverse
  field has a zero temperature phase transition driven solely by
  quantum fluctuations.  This quantum phase transition occuring at a
  critical transverse field strength has attracted much attention
  recently.  We report the progress that has been made via Monte Carlo
  simulations of the finite dimensional, short range model.}

\vspace*{0.6cm}
\normalsize\baselineskip=15pt
\setcounter{footnote}{0}
\renewcommand{\thefootnote}{\alph{footnote}}

\newcommand{\bc}{\begin{center}}
\newcommand{\ec}{\end{center}}
\newcommand{\ba}{\begin{array}}
\newcommand{\ea}{\end{array}}
\newcommand{\bt}{\begin{tabular}}
\newcommand{\et}{\end{tabular}}
\newcommand{\bi}{\begin{itemize}}
\newcommand{\ei}{\end{itemize}}
\newcommand{\be}{\begin{equation}}
\newcommand{\ee}{\end{equation}}

The most frequently discussed spin glass models are classical systems
for which quantum fluctuations can be neglected\cite{Binder86}. In
most cases this is correct, namely as long as $T_c>0$. The reason for
that is that critical fluctuations at the transition occur at a
frequency $\omega_c$ ($=\tau^{-1}$, where $\tau$ is the relaxation
time) with $\hbar\omega_c\ll k_B T$ since $\omega_c\to0$ as $T\to T_c$
due to critical slowing down. Hence any finite temperature will
destroy quantum coherence and the system will behave classically.
Very recently however, spin glasses began to enter the quantum
regime\cite{Sachdev94}.

\begin{figure}
\vspace*{13pt}
\psfig{file=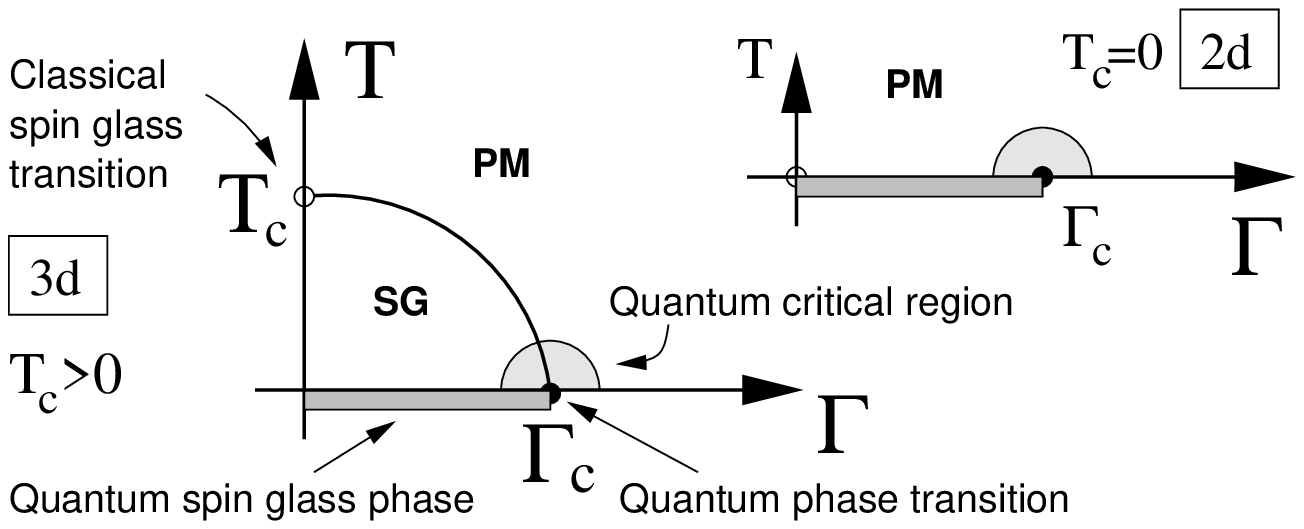,height=5.7cm}
\fcaption{
Typical phase digram of a three-- (left) and
two-- (right) dimensional quantum Ising
spin glass in a transverse field.}
\label{fig:qsg}
\end{figure}

The interesting theoretical question is: What are the effects of
quantum mechanics on the physics of strongly disordered systems at
zero temperature, where no heat bath is present and hopping over
energy barriers is replaced by tunneling through them quantum-mechanically?
The renewed interest in spin glasses in the quantum regime was kindled
by a series of recent experiments\cite{Wu91} on the dipolar Ising
magnet Li$_{x}$Ho$_{1-x}$YF$_{4}$, where $T_c$ was driven down to zero
by the application of a transverse magnetic field $\Gamma$ (see the
phase diagram depicted in figure \ref{fig:qsg}). Therefore it became
possible to study the {\it zero} temperature phase transition occuring
for critical transverse field strength $\Gamma_c$ by simply tuning the
external field. This quantum phase transition lies within a
different universality from the usually studied (classical) spin
glass transition at finite temperatures and it turns out that their
properties differ~significantly.

The above mentioned experiments can be described by the model
Hamiltonian of an Ising spin glass in a transverse field\cite{Wu91}
\be
H=-\sum_{\langle ij\rangle}J_{ij}\sigma_i^z\sigma_j^z-
\Gamma\sum_i\sigma_i^x
-h\sum_i \sigma_i^z\;,
\label{tisg}
\ee
where the $\sigma_i$ are Pauli spin matrices, $\Gamma$ is the strength
of the transverse field and $h$ is a longitudinal magnetic field used
to define magnetic susceptibilities but usually set to zero.
Obviously, for $\Gamma=0$ the quantum-mechanical Hamiltonian
(\ref{tisg}) is diagonal in the $z$-representaion of the spin
operators, which in this case can simply be replaced by their
eigenvalues $\pm1$ (after rescaling the couplings) giving exactly the
classical EA-model in $d$ dimensions: $H=-\sum_{\langle
  ij\rangle}J_{ij} S_i^z S_j^z$, where now $S_i=\pm1$ are Ising spin
variables. In this way the transverse field introduces quantum
mechanics into the spin glass problem and the value of $\Gamma$ tunes
the strength of the quantum fluctuations.  At zero temperature and
$\Gamma=0$ the system described by (\ref{tisg}) will be in its
uniquely determined ground state, which is identical to the classical
ground state of the EA-spin glass model.  In this case one has
$\langle\sigma_i^z\rangle=\pm1$ for all sites $i$ and therefore
$q_{EA}=[\langle\sigma_i^z\rangle^2]_{\rm av}=1$ , where
$\langle\cdots\rangle$ means the quantum-mechanical expectation value.

If we switch on the transverse field ($\Gamma>0$) the
Hamiltonian (\ref{tisg}) is not diagonal in the $z$-representation
any more and its ground state will be a superposition of the classical
ground state plus various excited states, which describes
the quantum-mechanical tunneling at zero temperature between the
local energy minima of the classical Hamiltonian. Furthermore
$|\langle\sigma_i^z\rangle|<1$ since the transverse
field tries to align the spins in the $x$-direction
and therefore $q_{EA}=[\langle\sigma_i^z\rangle^2]_{\rm av}<1$.
Increasing $\Gamma$ diminishes the EA-order parameter
$q_{EA}$ and for some critical value $\Gamma_c$ it will be
zero: $q_{EA}=0$ for $\Gamma\ge\Gamma_c$. This is the zero
temperature phase transition we are interested in and obviously we
cannot expect that its critical properties have anything in common
with the finite temperature classical spin glass transition.

In order to describe this zero-temperature transition one introduces a
quantity measuring the distance from the critical transverse field
strength (at $T=0$) $\delta=(\Gamma-\Gamma_c)/\Gamma_c$. If one
assumes a conventional second order phase transition one has
$q_{EA}\sim|\delta|^\beta$ and spatial correlations decay on a
characteristic length scale that diverges at the critical point as
usual: $\xi\sim|\delta|^{-\nu}$ and these exponents defined so far
would be sufficient to describe the static critical behavior of a
classical spin glass transition (provided hyperscaling holds).
However, at a zero temperature transition driven solely by quantum
fluctuations static and dynamic quantities are linked in such way that
the introduction of a characteristic time-scale (or inverse frequency)
is necessary: $\xi_\tau\sim\xi^z\sim|\delta|^{-z\nu}$, where $z$ is
the dynamical exponent. This will become more evident below, when we
consider an equivalent classical model.

Much work in the past has been devoted to the infinite range
model\cite{mft1} and the phase-diagram in the $\Gamma$-$T$-plane looks
similar to the one shown in figure \ref{fig:qsg}, i.e.\ for low enough
temperature $T$ and field $\Gamma$ one finds a transition line
separating a paramagnetic phase from a spin glass phase. Recent
work\cite{mft2} focused on the zero-temperature critical behavior and
calculated the critical exponents $\gamma=1/2$ (with multiplicative
logarithmic corrections), $\beta=1$ and $z\nu=1/2$.

In two and three dimensions, which are most relevant for the above
mentioned experiments, no analytical results are known. For this
reason extensive Monte Carlo simulations have been performed recently
in two dimensions by Rieger and Young\cite{Rieger94c} and in three
dimensions by Guo, Bhatt and Huse\cite{Guo94}. Usually the
investigation of quantum systems via Monte Carlo methods are hampered
by various deficiencies, the sign problem being the most notorious one
in this respect (see\cite{Linden92} for a review).  In studying the
Ising spin glass in a transverse field however, one can exploit the
fact that it can be mapped exactly onto a classical Ising model
described by a {\it real} Hamiltonian.  Using the Suzuki-Trotter
formula\cite{Suzuki76} one can easily show that the ground state
energy of the $d$--dimensional quantum mechanical model (\ref{tisg})
is equal to the free energy of a $(d+1)$--dimensional classical model,
where the extra dimension corresponds to imaginary time, i.e.
\be
-{\textstyle  E( T = 0) \over \textstyle L^d} =
\lim_{T\to 0} {\textstyle T \over \textstyle L^d} {\rm Tr\;} e^{-\beta H}
 = {\textstyle 1 \over \textstyle \Delta\tau}
{\textstyle 1 \over \textstyle L_\tau L^d} {\rm Tr\;}  e^{-{\cal S}}
\label{suzuki}
\ee
where the imaginary time direction has been divided into $L_\tau$ time
slices of width $\Delta\tau$ ($\Delta\tau L_\tau = \beta$), and the
effective classical action, $\cal S$, is given by
\be
{\cal S} = -\sum_{\tau}\sum_{\langle ij\rangle}
K_{ij} S_{i}(\tau) S_{j}(\tau)
-\sum_{\tau}\sum_i K S_{i}(\tau) S_{i}(\tau+1)
-\sum_{\tau}\sum_i H S_{i}(\tau)\;,
\label{class}
\ee
where the $S_i(\tau)=\pm1$ are classical Ising spins, the indices $i$
and $j$ run over the sites of the original $d$--dimensional lattice
and $\tau = 1,2,\ldots,L_\tau$ denotes a time slice. Moreover
it is $K_{ij} = \Delta\tau J_{ij}$, $H = \Delta\tau
h$ and $\exp(-2 K) = \tanh(\Delta\tau \Gamma)$. One has the {\em same}
random interactions in each time slice.  In order to fulfill the
second equality in (\ref{suzuki}) precisely, one has to perform the
limit $\Delta\tau\to 0$, which implies $K_{ij} \to 0$ and $K \to
\infty$. However, the universal properties of the phase transition are
expected to be independent of $\Delta\tau$ so we take $\Delta\tau = 1$
and set the standard deviation of the $K_{ij}$ to equal $K$. Thus $K$,
which physically sets the relative strength of the transverse field
and exchange terms in (\ref{tisg}), is like an inverse ``temperature''
for the effective classical model in (\ref{class}).

One sees that the (d+1)-dimensional classical model (\ref{class})
should order at low ``temperature'' (or coupling constant $K$) like a
spin glass in the $d$ spatial dimensions and ferromagnetically in the
imaginary time direction. From this one concludes the existence of two
different diverging length-scales in the classical model
(\ref{class}): one for the spatial (spin glass)-correlations, which is
$\xi$, and one for imaginary time (ferromagnetic) correlations, which
is $\xi_\tau$. Thus in the representation (\ref{class}) the link
between statics and dynamics in the original quantum model
(\ref{tisg}) becomes most obvious. Correspondingly, to analyze the
critical properties of the extremely anisotropic classical model
(\ref{class}) one has to take into account these two length scales via
anisotropic finite size scaling\cite{Binder89}.

\begin{figure}
\vspace*{13pt}
\psfig{file=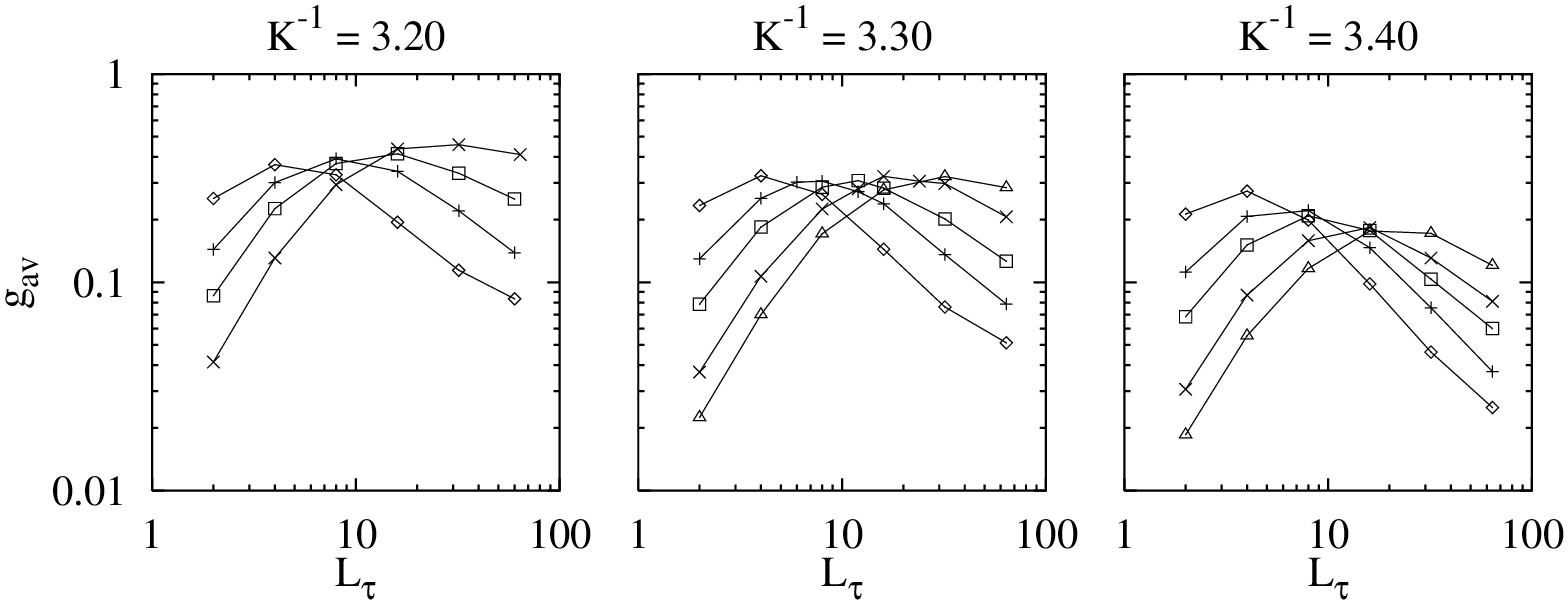,height=6.0cm}
\vskip-0.8cm
\fcaption{
The dimensionless cumulant $g_{\rm av}$ for the (2+1)-dimensional
classical model (\ref{class}) for three different values of the
coupling constant versus the system size $L_\tau$ in the imaginary
time direction. The system size in the space direction is
$L=4$ ($\diamond$), $6$ ($+$), $8$ ($\Box$), $12$ ($\times$) and
$16$ ($\triangle$). Since the maximum of $g_{\rm av}(L_\tau)$ is
roughly independent of $L$ at $K^{-1}\approx3.30$ one concludes that the
latter value is the critical coupling constant.}
\label{fig:g}
\end{figure}

Monte Carlo simulations of the classical model (\ref{class})
are straightforward but it turns out that sample-to-sample
fluctuations are significant, so one has to
do an extensive disorder average\cite{Rieger94c,Guo94}.
However, the finite size scaling analysis is complicated by
the existence of two diverging length scales
$\xi$ and $\xi_\tau$ and one has to deal with two independent
scaling variables: as usual $L/\xi$ and in addition the
shape (or aspect ratio) $L_\tau/L^z$ of the system\cite{Binder89}.
Thus, with the usual definition of a spin glass overlap
$Q=L^{-d}L_\tau^{-1}\sum_{i,\tau} S_i^a(\tau) S_i^b(\tau)$
($a$ and $b$ are two replicas of the same system)
for the classical system, the dimensionless combination of
moments of the order-parameter $g_{\rm av}$ obeys
\be
g_{\rm av}(K,L,L_\tau)=
0.5[3-\langle Q^4\rangle/\langle Q^2\rangle^2]_{\rm av}
\sim\tilde{g}_{\rm av}(\delta L^{1/\nu},L_\tau/L^z)\;.
\label{hypo}
\ee
Here $\langle\cdots\rangle$ means the thermal average and $[\cdots]_{\rm av}$
means the disorder average.
In isotropic systems one has $z=1$, which makes the aspect ratio
constant to one for the choice $L=L_\tau$ and in order to
determine the critical coupling $K_c$ one exploits the fact that
$g_{\rm av}(K,L,L)$ becomes independent of $L$ for $K=K_c$.
In the present case of a very anisotropic system
$z$ is not known a priori and one has to vary three different
system parameters to obtain an estimate for $K_c$ and $z$ (and other
exponents).  The following method\cite{Rieger94c,Guo94} enhances the
efficiency of such a search in a three-parameter space and also
produces reliable estimates for the quantities of interest: In the
limit $L_\tau\gg L^z$ the classical $(d+1)$-dimensional classical
system is quasi-one-dimensional, and in the limit $L_\tau\ll L^z$ the
system is quasi-$d$-dimensional and well above its
transition ``temperature'' in $d$ dimensions (which is even zero for
$d=2$). Therfore one has $\tilde{g}_{\rm av}(x,y)\to0$ both for $y\to0$ and
for $y\to\infty$.  Hence, for fixed $x$, $\tilde{g}_{\rm av}(x,y)$
must have a maximum for some value $y=y_{\rm max}(x)$. The value of
this maximum decreases with increasing $L$ in the disordered phase $K
< K_c$ (where $\delta = (K_c / K - 1) > 0$) and increases with
increasing $L$ in the ordered phase. This criterion can be used to
estimate the critical coupling, as exemplified in figure
(\ref{fig:g}). If one plots $g_{\rm av}(K_c,L,L_\tau)$ versus
$L_\tau/L^z$ with the correct choice for the dynamical exponent $z$,
one obtains a data-collaps for all system sizes $L$. Finally one
uses systems with fixed aspect ratio $L_\tau / L^z$ to determine
critical exponents via the usual one--parameter finite size scaling.

Various scaling predictions can be made if one supposes a conventional
second order phase transition to occur at some critical
``temperature'' $K_c$ for the classical model. Let us assume that we
are {\it at} the critical point $K=K_c$ and the aspect ratio stays
constant $L_\tau\propto L^z$. Then the order parameter scales with
system size as $q_{EA}=[|\langle Q\rangle|]_{\rm av}\sim
L^{-\beta/\nu}$ and the corresponding susceptibility $\chi_{SG}=L_\tau
L^d [\langle Q\rangle^2]_{\rm av}\sim L^{\gamma/\nu}$ with
$\gamma/\nu=d+z-2\beta/\nu$. The equal time correlation function in
the infinite system decays like $C(r)=[\langle
S_i(\tau)S_{i+r}(\tau)\rangle^2]_{\rm av} \sim r^{-(d+z-2+\eta)}$,
from which one obtains by integrating over $r$ the scaling relation
$2\beta/\nu=d+z-2+\eta$. The (imaginary)-time-dependent
autocorrelation function of the infinite system decays like $G(t) =
[\langle S_i(\tau)S_{i}(t+\tau)\rangle]_{\rm av}\sim t^{-(d + z - 2 +
  \eta)/2z}$ and by integrating over $t$ one gets the uniform
susceptibility $\chi_F(\omega=0) = \partial
[\langle\sigma_i^z\rangle]_{\rm av}/\partial h \sim L^{-\gamma_f/\nu}$
with $\gamma_f=z-\beta/\nu$. Thus $\chi_f$ will diverge if
$z>\beta/\nu$.

The nonlinear susceptibility
$\chi_{nl}=\partial^3[\langle\sigma_i^z\rangle]_{\rm av} / \partial
h^3 \sim L^{-\gamma_{nl}/\nu}$ is most relevant for the
experiments\cite{Wu91}. This quantity, also given by
$\chi_{nl}=[3\langle M^2\rangle^2 -\langle M^4\rangle]_{\rm av}/L^d$,
with $M=\sum_{i,\tau}S_i(\tau)$, can be shown\cite{Rieger94c} to scale
in the same way as $L^d L_\tau^3[\langle Q^2\rangle]_{av}$, which
yields the scaling relation $\gamma_{nl}=\nu(2-\eta+2z)$. The
nonlinear susceptibility of the infinite system obeys
$\chi_{nl}(T,\delta)\sim\delta^{-\gamma_{nl}}\tilde{\chi}_{nl}(
T/\delta^{z\nu}\Bigr)$ with $\tilde{\chi}_{nl}(x)\to const.\ $ for
$x\to0$ and $\tilde{\chi}_{nl}(x)\to x^{-\gamma_{nl}/z\nu}$ for
$x\to\infty$. By setting $\Gamma=\Gamma_c$ one gets $\chi_{nl}\sim
T^{-\gamma_{nl}/z\nu}$ and similarily $\chi_F\sim T^{-\gamma_f/z\nu}$.

In table \ref{tab:exp} we list the results obtained so far in various
dimensions. The symbol $div$.\ in the second column means that first
and higher derivatives of the magnetization diverge already in the
disordered phase. The word $finite$ means that in three dimensions and
for $d$ larger than the upper critical dimension the uniform
susceptibility does not diverge at the critical point. The results in
the second column\cite{Fisher92} were obtained within a
renormalization group calculation, those in columns 3 to 5
with Monte Carlo simulations\cite{Crisanti94,Rieger94c,Guo94}, column
6 shows the result of a Migdal-Kadanoff RNG calculation and the last
column depicts analytical results from mean-field theory\cite{mft2}.
Note that in the case $d=1$ we refer to the non-frustrated random
transverse Ising chain, where the order parameter is the magentization
instead of the spin glass overlap and the exponents listed refer to
the critical behavior of quantities that are defined accordingly.

\bc
\begin{table}[h]
  \tcaption{Summary of critical exponents characterizing the quantum
    phase transition of the transverse field Ising spin glass in
    finite dimensions}\label{tab:exp} \bc
\begin{tabular}{|c||c|c|c|c|c|c|}\hline

& d=1 RG\cite{Fisher92} & d=1 MC\cite{Crisanti94} &
  d=2 MC\cite{Rieger94c} & d=3 MC\cite{Guo94} & d=3
  MK\cite{Boechat94} & d$\ge$8 MFT\cite{mft2}\\
\hline

$z$ & $\infty$ & $\sim 1.7$ & $1.50\pm0.05 $ & $\sim 1.3$ & $\sim1.4$ & $2$\\

$\nu$ & $2 $ & $\sim1.0$ & $1.0\pm0.1$ & $\sim 0.8$ & $\sim0.87 $ & $1/4$\\

$\eta $ & $0.38\ldots$ & $0.40\pm0.03$ & $0.50\pm0.05$ & $\sim0.9$ & --- &
$2$\\

$\beta/\nu$ &  $0.19\ldots$ &$0.18\pm0.02$ & $1.0\pm0.1$ & $\sim1.6$ &
$\sim1.5$ & 1/2 \\

$\gamma$ & {\it div.}\ & $2.3\pm0.1$ & $\sim1.5$ & $\sim0.9$ & $\sim1.2$ &
1/2\\

$\gamma_f$ & {\it div.}\ & --- & $\sim0.5$ & {\it finite} & --- & {\it
finite}\\

$\gamma_{nl} $ & {\it div.}\ & --- & $\sim4.5$ & $\sim3.5$ & --- & --- \\

\hline

\end{tabular}
\ec
\end{table}
\ec

These results imply in three dimensions a strong divergence of the
nonlinear susceptibility when approaching $T=0$: $\chi_{nl}\sim
T^{-2.7}$ (since $\gamma_{nl}/\nu z\approx2.7$), with a power that is
very close to the one for classical Ising spin glasses
($\sim2.9$, ref.\cite{Binder86}). This is in striking contradiction to the
observation made in the experiments mentioned in the
introduction\cite{Wu91} and in order to clarify this discrepancy
further work is necessary.


\begin{thebibliography}{9}

\small

\bibitem{Binder86}
        K.~Binder, A.~P.~Young,
        Rev.\ Mod.\ Phys.\ {\bf 58}, 801 (1986).

\bibitem{Sachdev94}
        S.\ Sachdev, Physics World {\bf 7} (1994) 25.

\bibitem{Wu91}
        W. Wu, B. Ellmann, T. F. Rosenbaum, G. Aeppli, D. H. Reich,
        Phys. Rev. Lett. {\bf 67} (1991) 2076;
        W. Wu, D. Bitko, T. F. Rosenbaum, G. Aeppli,
        Phys. Rev. Lett. {\bf 71} (1993) 1919.

\bibitem{mft1}
        See Y. Y. Goldschmidt, P. Y. Lai,
        Phys. Rev. Lett. {\bf 64} (1990) 2567, and references therein.

\bibitem{mft2}
        J. Miller, D. Huse,
        Phys. Rev. Lett. {\bf 70} (1993) 3147;
        J. Ye, S. Sachdev, N. Read,
        Phys. Rev. Lett. {\bf 70} (1993) 4011;
        N. Read, S. Sachdev, J. Ye,
        preprint cond-mat/941203;
        J.\ W.\ Hartman, P.\ B.\ Weichman,
        Phys. Rev. Lett. {\bf 74} (1995) 4584.

\bibitem{Rieger94c}
        H. Rieger, A. P. Young,
        Phys.\ Rev.\ Lett.\ {\bf 72} (1994) 4141.

\bibitem{Guo94}
        M.~Guo, R.~N.~Bhatt, D. Huse,
        Phys.\ Rev.\ Lett.\ {\bf 72} (1994) 4137.

\bibitem{Linden92}
        W.\ von der Linden, Phys.\ Rep.\ {\bf 222} (1992) 53.

\bibitem{Suzuki76}
        M. Suzuki, Progr. Theor. Phys. {\bf 56} (1976) 1454.

\bibitem{Binder89}
        K. Binder, J. S. Wang, J. Stat. Phys. {\bf 55} (1989) 87.

\bibitem{Boechat94}
        M.\ A.\ Continentino, B.\ Boechat, R.\ R.\ dos Santos,
        Phys.\ Rev.\ B {\bf 50} (1994) 13528.

\bibitem{Fisher92}
        D. S. Fisher, Phys. Rev. Lett. {\bf 69} (1992) 534;
        Phys.\ Rev.\ B {\bf 51} (1995) 6411.

\bibitem{Crisanti94}
        A.\ Crisanti, H.\ Rieger,
        J.\ Stat.\ Phys.\ {\bf 77} (1994) 1087.


\end{thebibliography}
\end{document}